\documentclass[pra,preprint,superscriptaddress,floatfix,showpacs, 11pt]{revtex4-1}
\usepackage{hyperref}
\usepackage{amsmath,amsfonts,amssymb}
\usepackage{graphicx}
\usepackage{xcolor}
\usepackage{natbib}

\begin{document}
\title{Designing quantum dynamics in squeezed potentials}
\author{Rhombik Roy}
\email{rroy@campus.haifa.ac.il}
\affiliation{Department of Physics, University of Haifa, Haifa 3498838, Israel}
\affiliation{Haifa Research Center for Theoretical Physics and Astrophysics, University of Haifa,
Haifa 3498838, Israel}
\author{Ofir E. Alon}
\affiliation{Department of Physics, University of Haifa, Haifa 3498838, Israel}
\affiliation{Haifa Research Center for Theoretical Physics and Astrophysics, University of Haifa,
Haifa 3498838, Israel}

\begin{abstract} 
Geometry offers a versatile platform to control quantum transport. We investigate the ground-state properties and out-of-equilibrium dynamics of interacting bosons in a two-dimensional channel with two localized squeezes. The squeezes generate effective barriers along the longitudinal direction. These geometry-induced barriers modify the ground-state density distribution and lead to the emergence of many-body fragmentation. The squeezed channel also provides control over the transport dynamics, including the degree of self-trapping and the oscillation period. At the many-body level, the transport develops dynamical fragmentation, leading to the collapse of the density oscillations. 
We further demonstrate that introducing the two squeezes from opposite transverse directions generates oscillatory angular momentum during transport. The modified geometry also exhibits distinct many-body collapse behaviors associated with different patterns of fragmentation. Together, these results establish engineered squeezed confinement as a powerful tool for designing the ground-state properties and controlling the quantum transport, including angular-momentum generation and rich many-body transport dynamics.
\end{abstract}

\maketitle

\section{Introduction}\label{intro} 
Recent progress in ultracold atomic systems has provided greater flexibility in controlling their physical properties through engineered confinement geometries. Magnetic microtraps and atom chips first provided strongly confined waveguides and microscopic trapping structures for Bose–Einstein condensates~\cite{intro1}, with programmable atom-chip potentials providing further control over spatially varying confinement~\cite{intro4}. Optical techniques offer further flexibility through dynamically reconfigurable trapping potentials~\cite{intro2}. Spatial light modulators and digital micromirror devices allow greater precision in the design of trapping geometries~\cite{intro3}. These advances allow precise control over the shape, width, and dimensionality of matter-wave systems. Such control enables coherent transport through engineered structures~\cite{intro5} and opens new possibilities for atomtronic circuits~\cite{intro6}.

Different confinement geometries allow us to explore a variety of quantum phenomena in ultracold gases. For example, ring-shaped and toroidal traps have been used to explore persistent currents and superfluid transport~\cite{intro21,intro8}. 
Curved and shell-shaped geometries have attracted considerable interest for studying collective excitations, vortex dynamics, and the influence of curvature on superfluid phase transitions~\cite{intro7,intro23,intro29}. 
The experimental realization of ultracold atomic bubbles in microgravity has further advanced research on these systems~\cite{intro24}. 
Recent studies have explored the stability and dynamics of two-component Bose–Einstein condensates~\cite{intro25}, supersolidity and rotational effects~\cite{intro26}, as well as the role of shell-shaped confinement in collective excitations, thermodynamic properties, and topological superfluid phase transitions~\cite{intro27,intro28}. These studies show how the geometry of confinement influences the behavior of quantum gases and motivate further investigation of geometry-dependent quantum phenomena.

Studies have also shown that interparticle interactions can strongly modify resonant tunneling of Bose–Einstein condensates through localized barriers in mesoscopic waveguides~\cite{intro31,intro32}. Superfluid flow through narrow atomic channels have further highlighted the role of confinement in bosonic transport~\cite{intro10,intro33}. Changes in confinement can also affect the many-body state, giving rise to strong correlations and fragmentation even for weak interactions~\cite{intro11_sakmann,intro34}. More recently, bending a bosonic channel was shown to generate an effective barrier and modify tunneling, localization, and many-body correlations~\cite{rhombik_jcp}. 
These findings raise the intriguing question of whether similar control can be achieved by locally squeezing a straight channel, potentially opening up new avenues for manipulating quantum transport. A narrower region in a channel increases the transverse confinement and the zero-point energy, which acts as an effective barrier along the longitudinal motion. These geometry-induced barriers offer a way to control the bosonic transport without introducing additional longitudinal potentials.

In this work, we study interacting bosons in a two-dimensional channel with two localized squeezes along its length. We first consider a symmetric geometry, in which the channel narrows from both the upper and lower boundaries at each squeeze. We investigate how increasing the squeeze magnitude modifies the ground-state density distribution and leads to the emergence of many-body fragmentation. We then study the out-of-equilibrium dynamics of bosons initially confined between the squeezed regions and examine how the channel geometry controls the transport. At the many-body level, fragmentation gradually builds up, leading to collapse of the  oscillations. We further extend this control by introducing a modified geometry with two squeezes applied from opposite transverse directions. This breaks the reflection symmetry of the channel and enables the generation of oscillatory angular momentum. The modified geometry also exhibits distinct many-body collapse behaviors associated with different patterns of dynamical fragmentation.

The paper is organized as follows. Section~\ref{system} presents the design principles and numerical methods. Section~\ref{results1} discusses the engineering of ground-state properties through squeezing the channel. Section~\ref{results2} investigates the control of the transport and many-body dynamics in the symmetric squeezed potential. In Sec.~\ref{results3}, we extend our investigation to the modified squeezed potential and explore how breaking the reflection symmetry gives rise to angular-momentum dynamics and distinct many-body phenomena. Finally, Sec.~\ref{conclusion} summarizes the main findings. Appendix~\ref{apndx1} presents the  exact diagonalization analysis of the low-lying eigenstates and their overlaps with the initial state, which helps identify the parameter regimes for full transport. The supplementary material contains further information on numerical convergence of the calculations.

\section{Design principles and Methods}\label{system}
\begin{figure}[t]
    \centering
    \includegraphics[width=0.8\columnwidth]{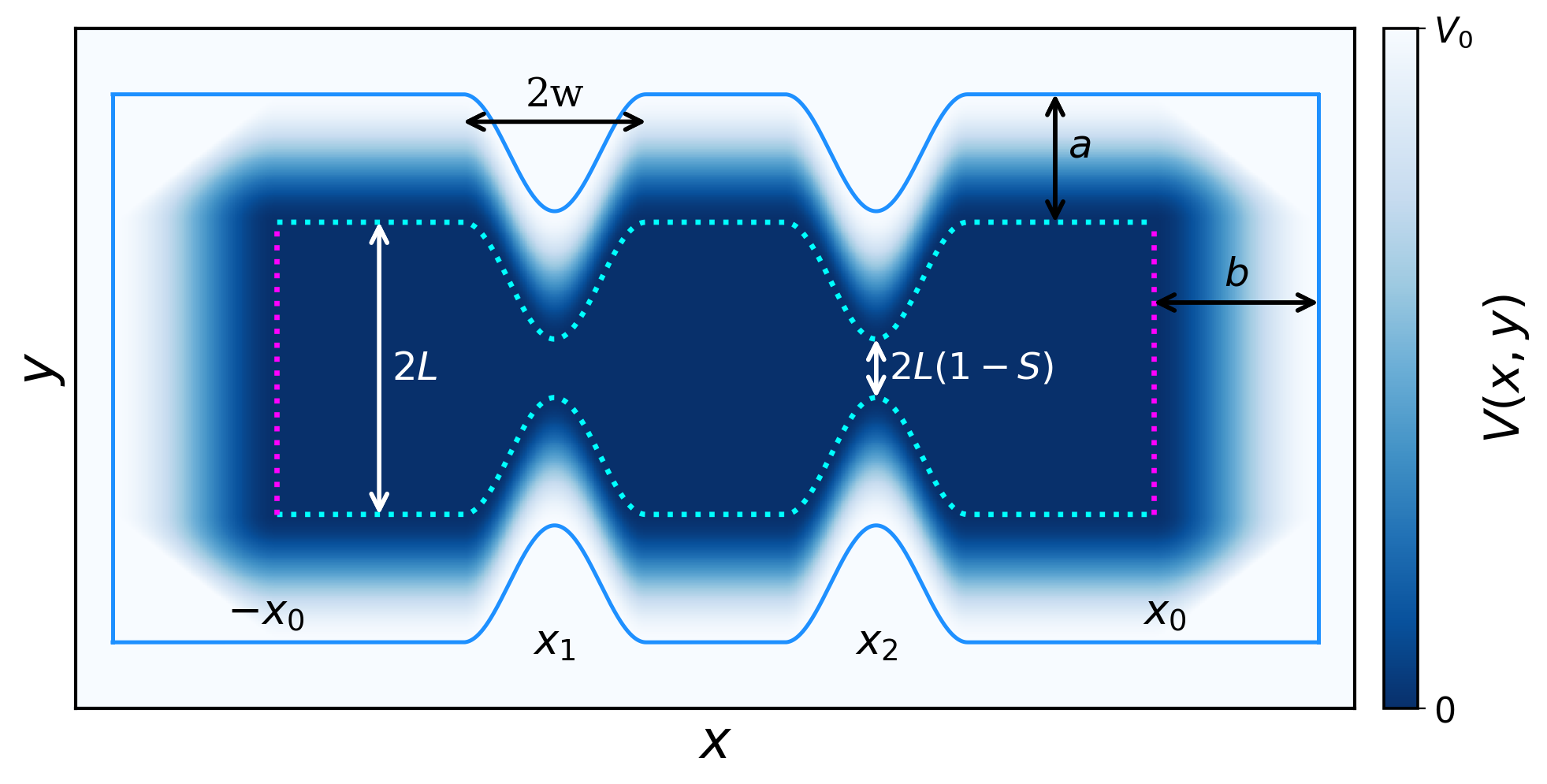}
    \caption{Schematic diagram of the two-dimensional squeezed-channel potential with two identical squeezes centered at $x_1$ and $x_2$. Each squeeze has the longitudinal width of $2w$. The dotted cyan curves mark the transverse boundaries of the zero-potential region, $y=\pm\ell(x)$. Beyond these boundaries, the potential rises smoothly from zero to $V_0$ over a transverse distance $a$. The outer soft-wall boundaries are shown by solid blue curves. Similarly, the dotted purple lines at $x=\pm x_0$ mark the longitudinal boundaries of the zero-potential region. Beyond these, the potential rises smoothly to $V_0$ over a longitudinal distance $b$. See the text for further details. All quantities shown are dimensionless. }
    \label{fig0}
\end{figure}

We consider a system of $N$ interacting bosons confined in a two-dimensional channel extending along the $x$ direction. As schematically illustrated in Fig.~\ref{fig0}, the channel contains two identical squeezes centered at $x_1$ and $x_2$, which locally narrow the transverse width of the channel. The engineered confinement allows the design of ground-state properties and control of quantum transport without introducing additional longitudinal potential barriers. We define the local half-width of the channel, $\ell(x)$, as the distance from the longitudinal centerline ($y=0$) to either transverse boundary (shown as cyan dotted line in Fig.~\ref{fig0}) at a given position $x$. Its spatial variation is given by
\begin{equation}
\ell(x)=L\left[1-S\sum_{j=1}^{2} \Theta\!\left(w-|x-x_j|\right)
\cos^{2}\!\left(\frac{\pi(x-x_j)}{2w}\right)\right],
\label{eq:sq1}
\end{equation}
where $L$ is the unperturbed channel half-width, $S$ is the squeeze magnitude, $2w$ is the longitudinal width of the squeezes, and $\Theta$ denotes the Heaviside step function. At the center of each squeeze, $x=x_j$, the channel reaches its minimum half-width, $\ell(x_j)=L(1-S)$.

The smooth confinement boundaries are described by the function
\begin{equation}
\mathcal{F}(u)=
\begin{cases}
0, & u\leq 0,\\[2pt]
\sin^{2}\left(\dfrac{\pi u}{2}\right), & 0<u<1,\\[4pt]
1, & u\geq 1.
\end{cases}
\label{eq:soft_wall}
\end{equation}
The transverse confinement is then written as
\begin{equation}
V_{\perp}(x,y)=V_0 \, \mathcal{F}\left(\frac{|y|-\ell(x)}{a}\right),
\label{eq:sq1t}
\end{equation}
where $a$ is the transverse soft-wall width and $V_0$ is the maximum potential height. The potential is zero inside the channel, $|y|\leq\ell(x)$, and gradually increases to $V_0$ over a transverse distance $a$.

To confine the bosons also along the longitudinal direction, the channel is closed by two identical smooth walls located symmetrically at $x=\pm x_0$. Using the same soft-wall function defined in Eq.\eqref{eq:soft_wall}, the longitudinal potential is written as
\begin{equation}
V_{\parallel}(x)=
\begin{cases}
0, & \vert x \vert\leq x_0,\\[2pt]
V_0 \,\mathcal{F} \left(\frac{|x|-x_0}{b} \right), & \vert x \vert > x_0,
\end{cases}
\label{eq:sq1l}
\end{equation}
where $b$ determines the width of the longitudinal boundaries. The potential vanishes for $|x|\leq x_0$ and smoothly increases to $V_0$ over a distance $b$. 
Combining Eqs.\eqref{eq:sq1t} and \eqref{eq:sq1l}, the full two-dimensional confining potential is expressed as
\begin{equation}
V(x,y) =\min\left[V_0, \, V_{\perp}(x,y)+V_{\parallel}(x) \right].
\label{eq:squeezed_potential1}
\end{equation}
This potential describes a reflection-symmetric channel with two localized squeezes and smooth boundaries in both the transverse and longitudinal directions, see Fig.~\ref{fig0}. 

The properties of the interacting bosonic system in the squeezed-channel potential are described by solving the time-dependent many-body Schr\"odinger equation, $ i\frac{\partial \Psi}{\partial t} = \hat{H} \Psi$, where $\hat{H}$ is the Hamiltonian, given by
\begin{equation} \label{hamiltonian}
\hat{H} = \sum_{j=1}^{N} \hat{h}(\mathbf{r}_j) + \sum_{j<k} \hat{W}(\mathbf{r}_j - \mathbf{r}_k).
\end{equation}
Here, $\hat{h}(\mathbf{r}_j)=-\frac{1}{2}\nabla_j^2+V(\mathbf{r}_j)$ is the single-particle Hamiltonian, with $\mathbf{r}_j=(x_j,y_j)$ denoting the position of the $j$-th boson. The second term accounts for the interactions between the bosons. The interparticle interaction is described by finite-range Gaussian potential, $\hat{W}(\mathbf{r}_j - \mathbf{r}_k) = \frac{\lambda_0}{2\pi \sigma^2} \exp\left[-\frac{(\mathbf{r}_j - \mathbf{r}_k)^2}{2\sigma^2}\right]$, where $\lambda_0$ is the interaction strength and $\sigma=0.25$~\cite{gaussian1,gaussian2,gaussian3}. The interaction parameter is defined as $\Lambda = \lambda_0 (N - 1)$. In the mean-field description, the ground-state and dynamics are governed by $\Lambda$. Thus, the mean-field properties remain unchanged with particle number as long as $\Lambda$ is fixed.

We employ the multiconfigurational time-dependent Hartree method for bosons (MCTDHB) to solve the schr\"odinger equation~\cite{MCTDHB1,MCTDHB2,MCTDHB11}. This method is the bosonic extension of the multiconfiguration time-dependent Hartree method, which has been extensively used in molecular quantum dynamics~\cite{1mctdh,2mctdh,3mctdh,4mctdh,5mctdh,6mctdh}. The $N$-boson wave function is expanded in bosonic permanents constructed from $M$ time-dependent single-particle orbitals, ${\phi_j(\mathbf r,t)}_{j=1}^{M}$, as
\begin{equation} 
\vert \Psi(t)\rangle = \sum_{n} C_{n}(t) \vert n;t\rangle , \qquad \sum_{j=1}^{M}n_j=N , 
\end{equation} 
where $n=(n_1,\ldots,n_M)$ denotes the occupation of the $M$ orbitals. MCTDHB allows both the coefficients $C_n(t)$ and the orbitals $\phi_j(\mathbf r,t)$ to evolve in time. Their coupled equations of motion are obtained variationally from the many-body Schr\"odinger equation~\cite{mctdhb_review}. This method captures many-body effects, such as correlations and fragmentation, beyond the mean-field description. The Gross–Pitaevskii equation is the special case of MCTDHB with single orbital $M=1$. Further details of the formalism and its numerical implementation can be found in Refs.~\cite{MCTDHB1,mctdhb_software1,mctdhb_software2}.

MCTDHB and the broader MCTDH-X framework have been applied to a variety of correlated quantum systems, including interacting bosons~\cite{fischer_budha,axel_spinor,mctdhb_exact3,sudip_asymmetric_dw,rhombik_epjd} and optical-lattice systems~\cite{paolo_mott,paolo_ol,Mistakidis_ol2,rhombik_pre,paolo_cz,rhombik_shubhro_pra}. Other applications include bosons in a cavity and engineered trapping potentials~\cite{paolo_cavity,paolo_cavity2,paolo_PhysRevA.98.053620,paolo_shake_potential}, rotating and non-inertial many-body dynamics~\cite{phantom_angular_axel,rhombik_scirep,paolo_rotation,beinke_rotation1,rhombik_acc,rhombik_infer_rot}, and quantum metrology~\cite{fischer_Metrology}. The corresponding fermionic formulation has also been used to investigate correlated fermionic systems~\cite{MCTDHB_fermion11,MCTDHB_fermion2,MCTDHB_fermion12,Elke_fermion,paolo_fermion}. The MCTDHB formalism has also been extended to the multilayer multiconfigurational time-dependent Hartree method for bosons (ML-MCTDHB)~\cite{ml-mctdhb1,ml-mctdhb2}.

\section{Ground-state engineering through channel squeezing}\label{results1}
\begin{figure}[hbpt!]
    \centering
    \includegraphics[width=0.6\columnwidth]{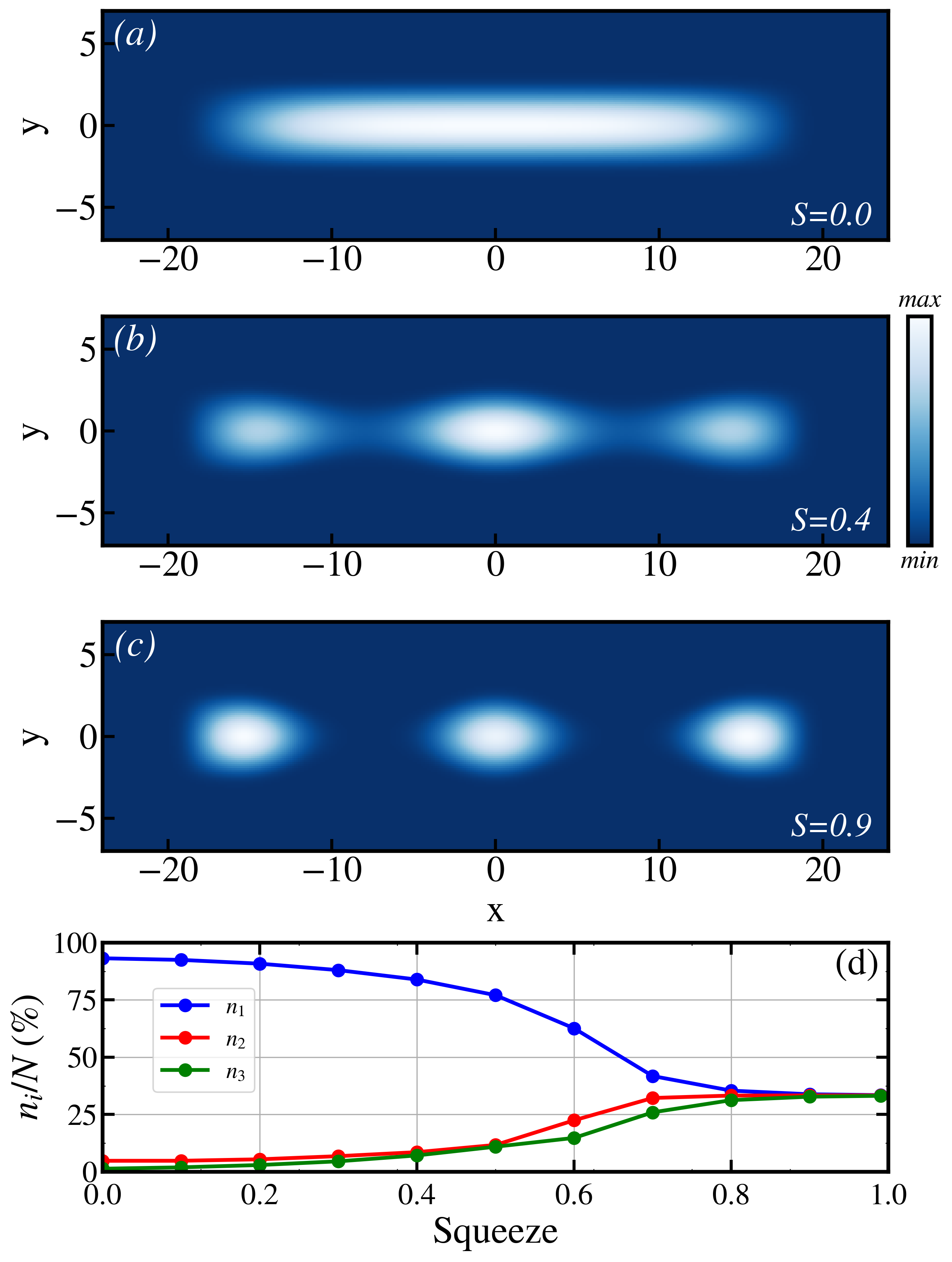}
    \caption{Localization of the ground-state density and fragmentation induced by squeezing the channel potential. Panels (a)-(c) show the mean-field one-body density for $S=0.0$, $0.4$, and $0.9$, respectively. The interaction parameter is fixed at $\Lambda=3.0$. The corresponding many-body densities for $N=6$ are nearly identical to the mean-field densities and are therefore not shown separately. With increasing $S$, the initially delocalized density cloud splits into three. Panel (d) shows the occupations of the three leading natural orbitals obtained from the many-body calculations as a function of $S$. The system evolves from a nearly condensed state toward a threefold fragmented ground state, with occupations $n_1 \approx 0.33$, $n_2 \approx 0.33$, and $n_3 \approx 0.33$ at $S=0.99$. All quantities shown are dimensionless.}
    \label{fig1}
\end{figure}

We begin by investigating how squeezing a two-dimensional channel modifies the ground state of interacting bosons. We employ both mean-field and many-body methods to study the ground-state properties. To this end, we vary the squeeze magnitude $S$ and examine the corresponding changes in the one-body density, as shown in Figs.~\ref{fig1}(a)-(c). The other parameters are kept fixed at $L=2.0$, $a=3.0$, $b=8.0$, $w=3.5$, $x_1=-7.8$, $x_2=+7.8$, and $x_0=18.0$, see Eqs.\eqref{eq:sq1}--\eqref{eq:squeezed_potential1} and Fig.~\ref{fig0}. 
The calculations are performed on a two-dimensional spatial domain $[-32,32]\times[-32,32]$, discretized using $256\times 256$ grid points. The convergence of the results with respect to the spatial grid resolution and the number of orbitals is demonstrated in the supplemental material. The interaction parameter is fixed at $\Lambda=\lambda_0(N-1)=3.0$, where $\lambda_0$ denotes the two-body interaction strength. For the many-body calculations, we consider $N=6$ interacting bosons, corresponding to $\lambda_0=0.6$. The choice of the small, finite particle number allows us to highlight the role of many-body effects in the ground state. The potential height is set to $V_0=10.0$ throughout this work.

Figures~\ref{fig1}(a)-(c) display the ground-state one-body densities obtained from the mean-field calculation. The corresponding many-body densities are the same as the mean-field results and are therefore not shown separately. In the straight channel, $S=0$, the bosons form a single delocalized density cloud extending along the longitudinal direction [Fig.~\ref{fig1}(a)]. Increasing the squeeze strength  gradually splits the density into three spatial regions. This separation becomes visible at $S=0.4$ [Fig.~\ref{fig1}(b)] and for stronger squeezing, $S=0.9$, the density forms three clearly separated lobes [Fig.~\ref{fig1}(c)].
The formation of the three localized clouds can be understood from the local squeezing of the channel. As the squeeze magnitude $S$ increases, the transverse confinement becomes stronger around $x_1$ and $x_2$. This raises the transverse zero-point energy in these regions, which acts as an effective barrier along the longitudinal direction~\cite{waveguide,waveguide2}. As a result, the coupling between the central and outer regions becomes weaker, gradually separating the density into three clouds. Importantly, the three-cloud structure emerges entirely from the squeezing of the channel, without introducing any additional longitudinal potential barriers.

Although the mean-field and many-body analysis yield nearly identical one-body densities, the latter captures correlations and fragmentation absent in the mean-field description. To reveal these many-body effects, we examine the occupations in the natural orbital. The natural orbitals are obtained by diagonalizing the reduced one-body density matrix. The corresponding eigenvalues, known as the occupations in the natural orbitals, give the number of particles occupying each orbital~\cite{MCTDHB1}. In a fully condensed state, all bosons occupy a single natural orbital. When two or more natural orbitals become macroscopically occupied, the system is said to be fragmented~\cite{fragmentation1,fragmentation2,fragmentation3}.

Figure~\ref{fig1}(d) shows how the occupations of the natural orbitals evolve with increasing squeeze magnitude, revealing the gradual emergence of many-body fragmentation in the ground state. For weak squeezing, the first natural orbital has high occupation, and the contributions from the second and third orbitals remain small. As $S$ increases, $n_1$ decreases and $n_2$ and $n_3$ increase. For sufficiently strong squeezes the resulting geometry-induced barriers lead to a threefold fragmented ground state. At $S=0.99$, the three leading natural occupations become approximately equal, $n_1\approx n_2\approx n_3\approx 1/3$, indicating a nearly balanced threefold fragmented state. For the $N=6$ bosons considered in the many-body calculations, this corresponds to an occupation of two bosons in each of the three natural orbitals. Thus, the strong squeeze provides a direct realization of the engineered $\vert 2,2,2\rangle$ state. These results demonstrate that varying the squeeze magnitude controls not only the spatial distribution of the bosons but also the fragmentation of the many-body ground state. Further modifications of the squeeze positions, widths, and symmetry may offer additional control for designing different fragmented ground states.

\section{Tunneling and transport in the squeezed potential}\label{results2}
Having established how squeezing a channel with homogeneous width modifies the ground-state structure, we now investigate the out-of-equilibrium dynamics through the channel. Initially, all bosons are confined to the central region. To prepare the initial state, we confine all bosons to the central region by placing the left and right longitudinal boundaries at $x_1$ and $x_2$, respectively. The initial cloud is therefore restricted to the region $x_1<x<x_2$. At $t=0$, the boundaries are shifted back to $\pm x_0$, allowing the cloud to tunnel into the two outer regions. The interaction parameter is fixed at $\Lambda=0.2$ throughout the dynamics. 

To quantify the transport, we introduce the survival probability in the central region,
\begin{equation}
\mathcal{P}_{M}(t)=\frac{1}{N}\int_{x_1}^{x_2} dx \int_{-\infty}^{\infty}dy\, \rho(x,y;t),
\label{eq:survival}
\end{equation}
where $\rho(x,y;t)$ is the one-body density. Initially, all bosons are confined to the central region, giving $\mathcal{P}_{M}(0)=1$. The numerical calculations are performed using $128\times128$  grid points and the convergence with respect to the grid resolution is presented in the supplemental material.

\subsection{Controlling the tunneling dynamics}
\begin{figure}[t]
    \centering
    \includegraphics[width=0.8\columnwidth]{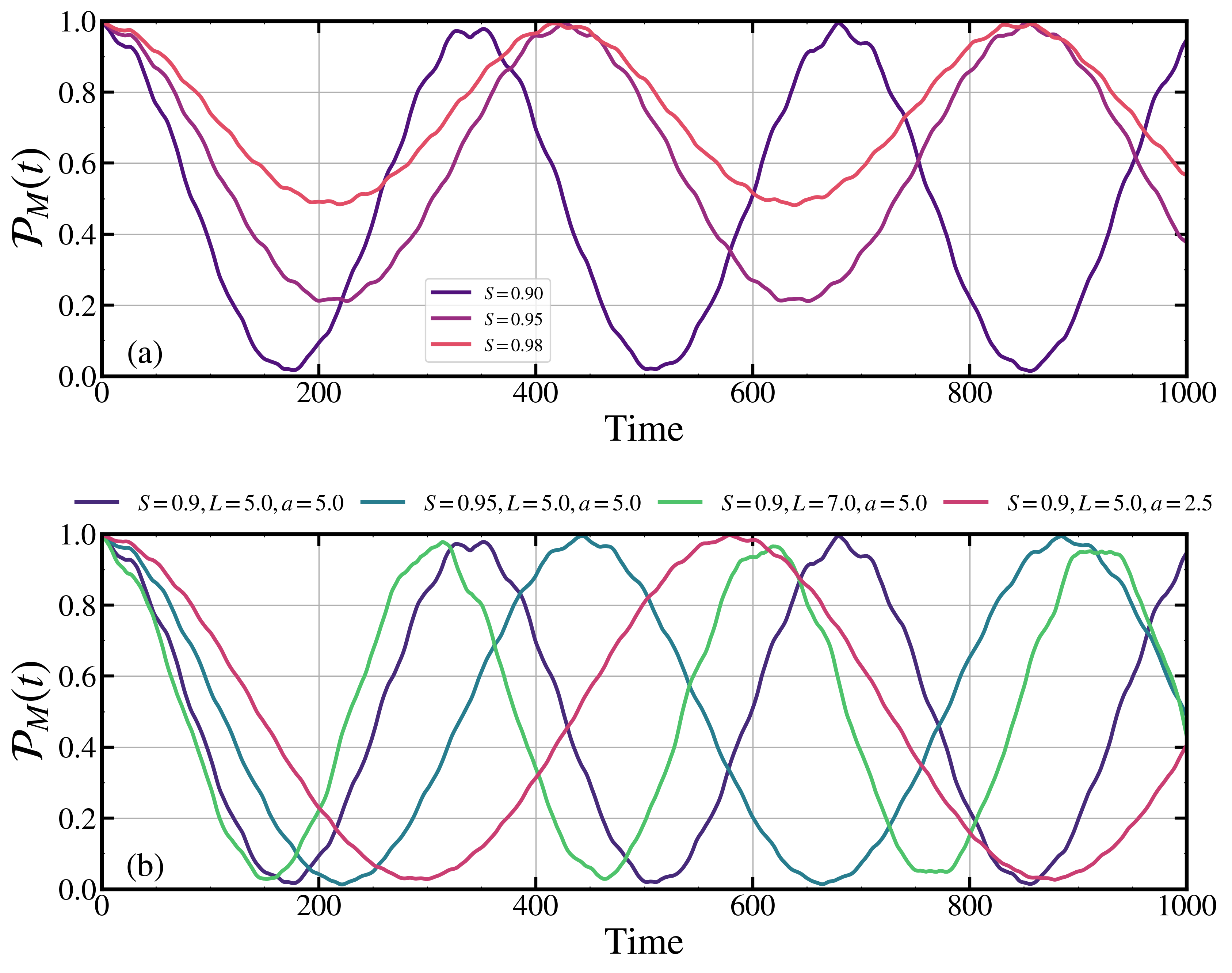}
    \caption{Transport in the squeezed potential, characterized by the survival probability $\mathcal{P}_{M}(t)$ in the central region. (a)  Survival probability for different squeeze magnitudes, $S=0.90$, $0.95$, and $0.98$, showing the transition from nearly complete tunneling to partial self-trapping. All other geometric parameters are fixed at $L=5.0$, $a=5.0$, $b=5.0$, $w=3.5$, $x_1=-4.0$, $x_2=+4.0$ and $x_0=10.48$, see Eqs.\eqref{eq:sq1}--\eqref{eq:squeezed_potential1} and Fig.~\ref{fig0}. (b) Control of the tunneling time period by varying $S$, $L$, and $a$ independently. For each parameter set, $x_0$ is adjusted to remain in the full-transfer regime. See text for more details. All quantities shown are dimensionless.}
    \label{fig2}
\end{figure}

Figure~\ref{fig2}(a) shows how the squeeze magnitude $S$ controls the amount of tunneling between the central and outer regions. The other parameters are fixed at $L=5.0$, $a=b=5.0$, $w=3.5$, $x_1=-4.0$, $x_2=+4.0$, and $x_0=10.48$, see Eqs.\eqref{eq:sq1}--\eqref{eq:squeezed_potential1} and Fig.~\ref{fig0}. Tunneling refers to the transfer of bosons between the central and outer regions through the barriers created by channel squeezing.  
The period of the tunneling oscillations corresponds to the time required for the bosons to tunnel from the central region to the outer regions and back. Self-trapping, on the other hand, occurs when a fraction of the bosons remains localized in the central region rather than participating in the tunneling process. Both the tunneling period and the degree of self-trapping can be extracted from the time evolution of the survival probability in the central region, $\mathcal{P}_M(t)$. In this subsection, we focus on the mean-field results, and the many-body dynamics are discussed in the following subsection. 
For $S=0.90$, the survival probability $\mathcal{P}_{M}(t)$ exhibits complete tunneling of the bosons from the central region to the two outer regions and back. Increasing the squeeze magnitude to $S=0.95$ leads to partial self-trapping, which becomes more pronounced at $S=0.98$. In the latter case, approximately half of the bosons remain confined to the central region throughout the time evolution, as evidenced by the oscillations of $\mathcal{P}_M(t)$ between $0.5$ and $1.0$. These results demonstrate how increasing the squeeze magnitude gradually suppresses tunneling and enhances self-trapping. Enlarging $S$ narrows the squeezed regions, which thereby increases the effective barriers and reduces the coupling between the central and outer regions. Therefore, the squeeze magnitude acts as a direct control parameter for the tunneling amplitude. Furthermore, due to the symmetry of the potential, particles tunnel equally from the central region into the two outer regions. 

The channel geometry also allows us to control the tunneling time period even in the full tunneling regime. Figure~\ref{fig2}(b) shows the survival probability for different parameter sets, all exhibiting complete tunneling but with different time periods.  Starting from the reference geometry $S=0.9$, $L=5.0$, and $a=5.0$, the tunneling time can be modified by changing the squeeze magnitude $S$, the channel half-width $L$, or the transverse confinement parameter $a$, see Eqs.\eqref{eq:sq1}--\eqref{eq:squeezed_potential1} and Fig.~\ref{fig0}. Enlarging the squeeze from $0.9$ to $0.95$ increases the time period. In contrast, increasing the channel half-width from $L=5.0$ to $L=7.0$ leads to a faster tunneling, and reducing $a$ from $5.0$ to $2.5$ considerably increases the time period. Thus, by adjusting the confinement parameters, both the self-trapping fraction and the tunneling period can be controlled independently.

To determine the parameter sets that yield the complete tunneling, we perform the exact-diagonalization analysis of the initial and post-quench Hamiltonians, as detailed in Appendix~\ref{apndx1}. By projecting the initially localized state onto the eigenstates of the post-quench potential, we adjust $x_0$ to obtain an overlap calculation that favors full tunneling; see Appendix~\ref{apndx1}. The tunneling dynamics are mainly governed by the first and third eigenstates of the post-quench Hamiltonian, with corresponding energies $E_1$ and $E_3$. Their energy splitting gives an estimate of the tunneling period, $T_{\rm ED}=2\pi/(E_3-E_1)$, which closely agrees with the numerical results with weak interaction. 

During the time evolution, the expectation values $\langle \hat{X}\rangle$, $\langle \hat{Y}\rangle$, $\langle \hat{P}_X\rangle$, $\langle \hat{P}_Y\rangle$, and $\langle \hat{L}_Z\rangle$ remain zero. This follows from the reflection symmetries of the channel and the symmetrically prepared initial state. Although these expectation values vanish, the tunneling dynamics produce time-dependent fluctuations of the corresponding observables. The dynamics of their spatial, momentum, and angular-momentum variances are presented and discussed in the supplemental material.

The above analysis demonstrates that the squeezed channel provides control over both the tunneling timescale and the fraction of self trapping. At the mean-field level, varying the squeeze magnitude allows us to control the transfer, whereas adjusting the other confinement parameters allows us to tune the tunneling period. We next investigate how this tunneling behavior changes at the many-body level.

\subsection{Many-body properties}\label{manybody1}
\begin{figure}[hpbt!]
    \centering
    \includegraphics[width=0.85\columnwidth]{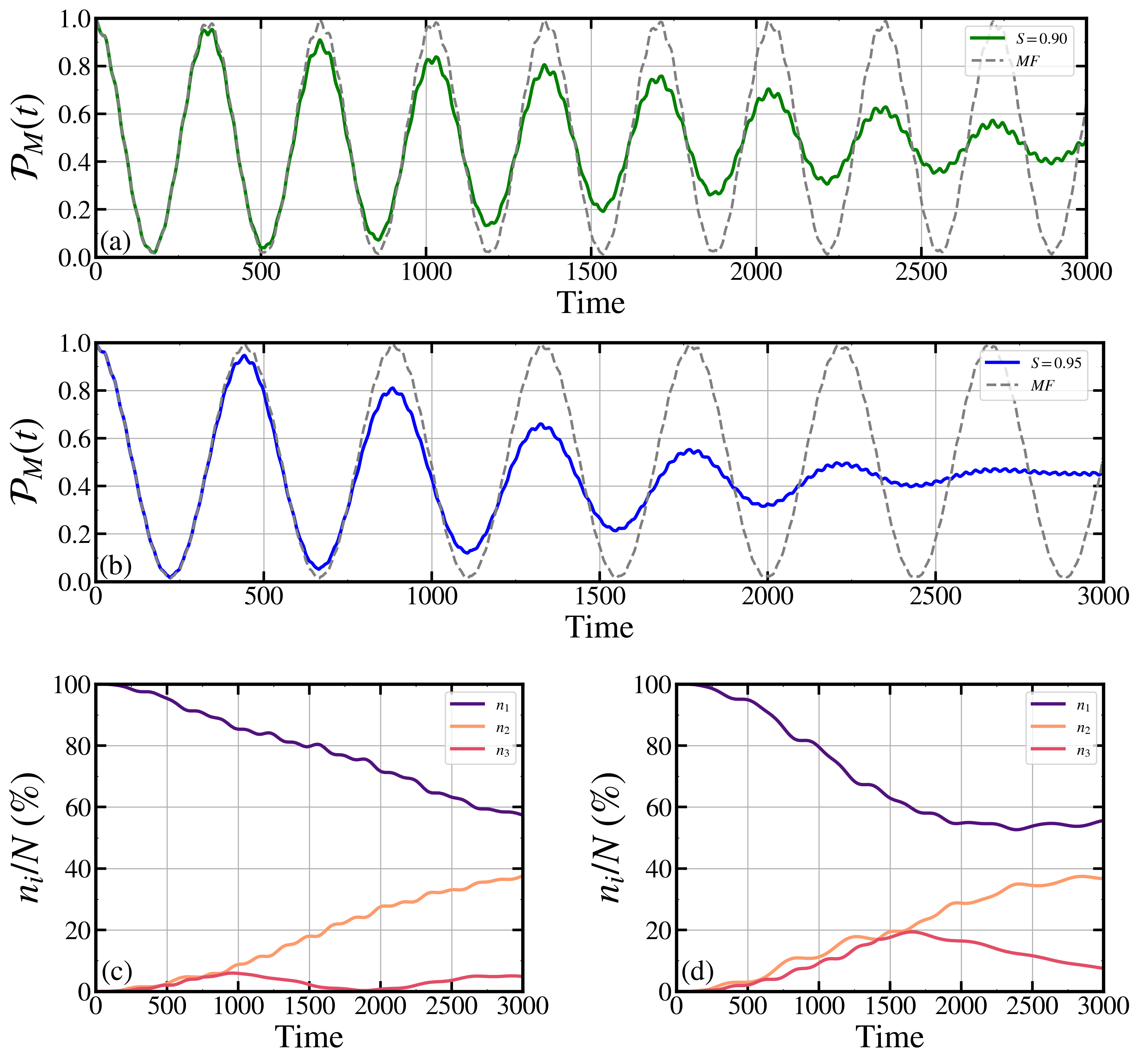}
    \caption{Many-body dynamics in the symmetric squeezed channel. (a,b) Survival probability in the central region, $\mathcal{P}_{M}(t)$, for $S=0.9$ and $S=0.95$, respectively. The corresponding mean-field (MF) results are shown as dashed gray lines. The many-body oscillations gradually decrease in amplitude, tending toward a collapse around $\mathcal{P}_{M} (t)\approx 0.5$. The higher squeeze magnitude exhibits faster collapse of the density oscillations. (c,d) Time evolution of the occupations in the first three natural orbitals for $S=0.9$ and $S=0.95$, respectively. The first two natural orbitals have a major contribution in the long-time dynamics. All quantities shown are dimensionless.}
    \label{fig3}
\end{figure}

We now investigate the transport in the symmetric squeezed channel at the many-body level of theory. The many-body calculations are performed for $N=10$ bosons using $M=6$ time-adaptive orbitals. The convergence of the results is presented in the supplemental material. Figures~\ref{fig3}(a) and (b) present the many-body survival probability $\mathcal{P}_M(t)$ for $S=0.9$ and $S=0.95$, respectively, alongside the corresponding mean-field results for comparison. All other parameters are fixed at $L=5.0$, $a=b=5.0$, $w=3.5$, $x_1=-4.0$ and $x_2=+4.0$. For both squeeze magnitudes, the mean-field and many-body dynamics agree closely at short times but gradually deviate as time evolves. While the mean-field dynamics exhibit full oscillations with constant amplitude, the many-body oscillations gradually decay and eventually collapse to $\mathcal{P}_M(t)\approx 0.5$. This corresponds to approximately half of the bosons occupying the middle region, with the remaining half equally distributed between the two outer regions. Comparing the two squeeze magnitudes, the overall tunneling behavior remains qualitatively similar in both descriptions, although the collapse of the many-body oscillations occurs more quickly for the larger squeeze magnitude.

To understand the collapse of the many-body oscillations, we examine the time evolution of the three macroscopically occupied natural orbitals. The results are shown in Figs.~\ref{fig3}(c) and (d) for $S=0.9$ and $S=0.95$, respectively. The occupations of the lower three orbitals are presented in the supplemental material where the convergence is discussed. For the symmetric squeezed potential, the qualitative feature for the two cases ($S=0.9$ and $S=0.95$) are same. Initially, the system is highly condensed, with $n_1\approx 1$. As time progresses, the occupation of the first natural orbital decreases and the occupation of the second natural orbital gradually increases, showing the buildup of fragmentation. The occupation of the third natural orbital also increases initially but remains below $10\%$ for $S=0.9$ and $20\%$ for $S=0.95$ throughout the evolution. At longer time, the first two natural orbitals remain dominant. This particular redistribution of natural occupations is associated with the gradual collapse of the survival probability oscillations toward $\mathcal{P}_{M}(t)\approx 0.5$, as observed in Figs.~\ref{fig3}(a) and (b). This is consistent with the fragmentation-induced damping reported in previous studies of many-body tunneling~\cite{anal_2022_dw,anal_scipost}.

In summary, the two-squeeze geometry provides control over both the amount of tunneling and its timescale. At the many-body level, the tunneling oscillations gradually collapse as fragmentation develops. Here, the first two natural orbitals remain dominant in the long-time dynamics. The reflection symmetry of the confinement, however, prevents the generation of net angular momentum, $\langle \hat{L}_Z\rangle=0$. We therefore explore how the geometry can be designed to generate angular momentum and introduce additional features in the many-body dynamics.

\section{Beyond symmetric squeezing}\label{results3}
Having established the control over particle transport in the symmetric squeezed channel, we now explore whether this control can be extended to angular-momentum dynamics. Angular-momentum dynamics are of particular interest in rotating Bose gases, including vortex formation and rotational excitations~\cite{connect2_fetter}. Controlled angular-momentum states are also closely connected with superfluid circulation and persistent currents, which have been observed experimentally in trapped Bose-Einstein condensates~\cite{connect1}. This naturally raises the question of whether the confinement can be engineered so that tunneling itself generates a finite angular momentum. In the following, we show that a suitable modification of the geometry can produce oscillatory angular-momentum dynamics.

\begin{figure}[t]
    \centering
    \includegraphics[width=0.8\columnwidth]{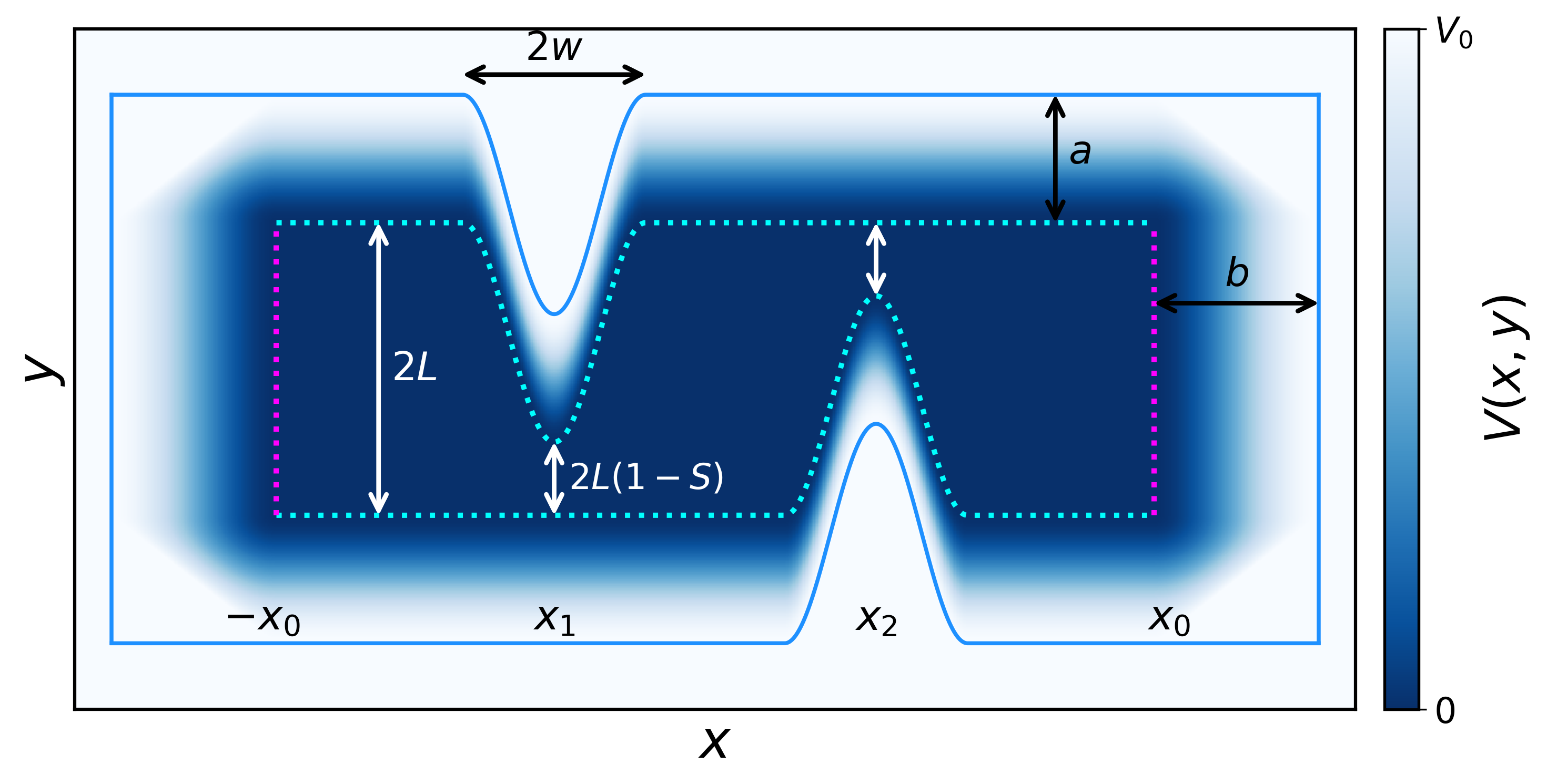}
    \caption{Schematic diagram of the modified squeezed potential with broken reflection symmetry. The left squeeze at $x_1$ is introduced from the upper boundary, while the right squeezing at $x_2$ is introduced from the lower boundary. The channel width is $2L$, with transverse and longitudinal soft-wall widths $a$ and $b$, respectively. Each squeeze has a longitudinal width $2w$, and its magnitude is controlled by $S$. See the text for further details. All quantities shown are dimensionless.}
    \label{fig4}
\end{figure}

In the modified geometry, schematically illustrated in Fig.~\ref{fig4}, the two squeezes are introduced from opposite transverse directions, thereby breaking the reflection symmetry of the channel. The left squeeze is created by moving the upper boundary downward, while the right squeeze is created by moving the lower boundary upward. We define the corresponding localized squeeze boundaries as
\begin{equation}
f_j(x)= \Theta\left(w-|x-x_j|\right) \cos^{2}\left[\frac{\pi(x-x_j)}{2w}\right], \qquad j=1,2,
\label{eq:sq2}
\end{equation}
where $x_1$ and $x_2$ denote the centers of the two squeezes and $2w$ is their longitudinal width. The upper and lower boundaries of the channel are then given by
\begin{equation}
\ell_{\rm t}(x)=L\left[1-2S f_1(x)\right],
\qquad
\ell_{\rm b}(x)=-L\left[1-2S f_2(x)\right],
\label{eq:sq22}
\end{equation}
where $L$ is the unperturbed half-width of the channel and $S$ is the squeeze magnitude. Thus, at $x=x_1$ only the upper boundary is modified, whereas at $x=x_2$ only the lower boundary is modified, as shown by the dotted cyan line in Fig.~\ref{fig4}.

Using the smooth-wall function $\mathcal{F}(u)$ defined in Eq.\eqref{eq:soft_wall}, the transverse confinement is written as
\begin{equation}
V_{\perp}(x,y)=
\begin{cases}
V_0\,\mathcal{F}\left( \frac{y-\ell_{\rm t}(x)}{a} \right), & >\ell_{\rm t}(x),\\[6pt]
0, & \ell_{\rm b}(x)\leq y\leq\ell_{\rm t}(x),\\[6pt]
V_0\,\mathcal{F}\left( \frac{\ell_{\rm b}(x)-y}{a} \right), &  <\ell_{\rm b}(x),
\end{cases}
\label{eq:sq2t}
\end{equation}
where $a$ is the transverse soft-wall width and $V_0$ is the maximum height of the potential. The potential vanishes inside the channel and rises sinusoidally from $0$ to $V_0$ over a distance $a$. The solid blue lines in Fig.~\ref{fig4} mark the positions at which the confinement potential reaches $V_0$. Thus, the potential creates two squeezes introduced from opposite transverse directions, each with a minimum channel width of $2L(1-S)$. 
The channel is closed in the longitudinal direction by two smooth walls located symmetrically at $x=\pm x_0$ and have the form
\begin{equation}
V_{\parallel}(x)=
\begin{cases}
0, & \vert x \vert\leq x_0,\\[2pt]
V_0 \,\mathcal{F} \left(\frac{|x|-x_0}{b} \right), & \vert x \vert > x_0,
\end{cases}
\label{eq:sq2l}
\end{equation}
where $b$ is the longitudinal soft-wall width. Hence, in the longitudinal direction, $V_{\parallel}(x)=0$ for $|x|\leq x_0$, increases smoothly over the region $x_0<|x|<x_0+b$, and it reaches $V_0$ for $|x|\geq x_0+b$.

The complete two-dimensional confinement is therefore written as
\begin{equation}
V(x,y)=\min\left[V_0,\,V_{\perp}(x,y)+V_{\parallel}(x)\right].
\label{eq:squeezed_potential2}
\end{equation}
Note that we use the same notation in Eqs.~\eqref{eq:squeezed_potential1} and \eqref{eq:squeezed_potential2} to avoid introducing additional parameters. The modified squeezed potential preserves rotational symmetry but breaks reflection symmetry, in contrast to the geometry shown in Fig.~\ref{fig0}, which possesses both symmetries.

\subsection{Generation and transport of angular momentum}
We now examine the dynamics in the modified squeezed channel introduced above [Fig.~\ref{fig4}]. The preparation protocol is kept the same as in the reflection symmetric potential: the bosons are initially localized in the central region between $x_1$ and $x_2$. At $t=0$, the bosons are allowed to tunnel toward the two outer regions. We keep the other geometric parameters fixed at $L=6.0$, $S=0.90$, $w=3.5$, $x_1=-4.0$, $x_2=4.0$, and $b=5.0$, and vary only the transverse confinement $a$. For each value of $a$, the longitudinal boundary position $\pm x_0$ is adjusted to observe complete tunneling. 
As discussed in Appendix~\ref{apndx1}, the full-tunneling can be identified from the exact-diagonalization analysis. In contrast to the reflection-symmetric potential, where equal overlaps of the initial state with the first and third eigenstates favor full tunneling, the present geometry exhibits full tunneling when the corresponding overlaps are approximately $60\%$ and $40\%$, respectively. This difference originates from the different spatial structure of the eigenstates in the transversely asymmetric channel (see Appendix~\ref{apndx1}). The period calculated from the exact diagonalization, $T_{\rm ED}=\frac{2\pi}{E_3-E_1}$, closely agrees with the oscillation period observed in the numerical simulations with weak interaction, where $E_1$ and $E_3$ denote the ground-state and second-excited-state energies, respectively. In this subsection, we focus on the mean-field dynamics. The corresponding many-body dynamics are discussed in the following subsection.

\begin{figure}[t]
    \centering
    \includegraphics[width=0.8\columnwidth]{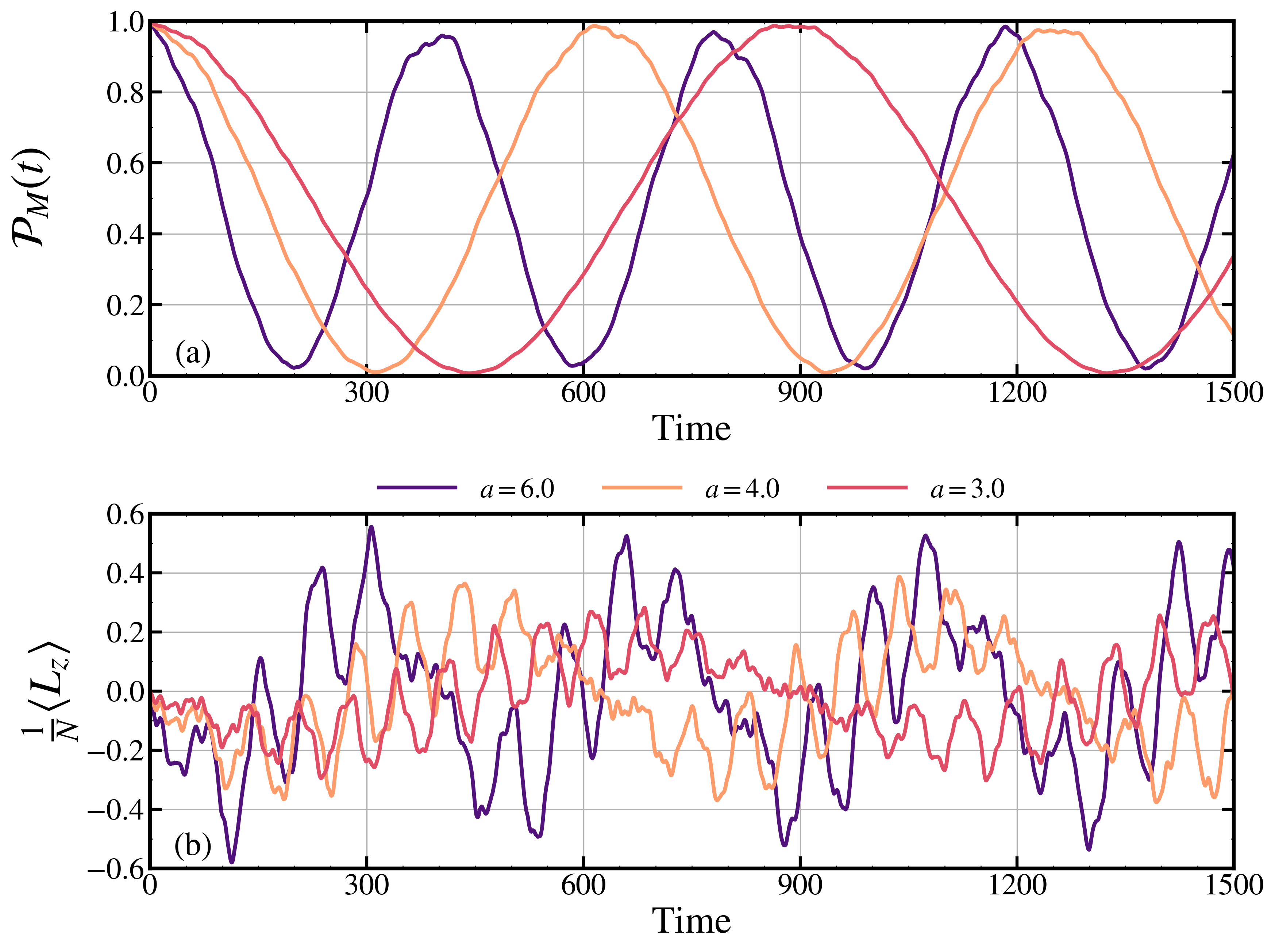}
    \caption{Survival probability and angular-momentum dynamics in the modified squeezed potential with broken reflection symmetry (see Fig.~\ref{fig4}) for different transverse confinement parameters. (a) Survival probability $\mathcal{P}_{M}(t)$ for transverse confinement parameters $a=3.0$, $4.0$, and $6.0$.  Increasing $a$ shortens the tunneling time period. (b) Corresponding angular momentum per particle, $\langle \hat{L}_Z\rangle/N$. The broken reflection symmetry gives rise to oscillatory angular momentum, with the amplitude of the oscillations increasing with $a$. See the text for further discussion. All quantities shown are dimensionless.}
    \label{fig5}
\end{figure}

Figure~\ref{fig5}(a) shows the survival probability $\mathcal{P}_{M}(t)$ in the modified squeezed potential for three different sharpness of the transverse confinement. The tunneling period is highly sensitive to the sharpness of the transverse confinement; see Fig.~\ref{fig4}. As $a$ increases from $3.0$ to $6.0$, the oscillations become faster. Therefore, the modified squeezed geometry also provides control to the tunneling period. Similar control can also be achieved by varying other confinement parameters (not shown here). 
Figure~\ref{fig5}(b) shows the average angular momentum per particle, $\langle \hat{L}_Z\rangle/N$, for the same three values of $a$. The modified geometry gives rise to a finite, oscillatory angular-momentum dynamics. As the transverse soft-wall width $a$ increases, the oscillation amplitude increases, demonstrating direct control of the angular-momentum dynamics through the transverse confinement.

The finite angular momentum originates directly from the pattern of flow of the bosons through the potential. At $x_1$, the channel becomes narrow from the upper side, so the bosons mainly pass through the lower part of the channel. At $x_2$, the geometry is reversed, and the particles mainly pass through the upper part. The two outgoing streams therefore follow spatially displaced paths and break the mirror symmetry with respect to the longitudinal axis.
As $\hat{L}_z=\sum_{i=1}^N\hat{x}_i\hat{p}_{y,i}-\hat{y}_i\hat{p}_{x,i}$, particles moving out to the left have $p_x<0$ at $y<0$, while particles moving out to the right have $p_x>0$ at $y>0$. Hence, the $-\hat{y}\hat{p}_x$ contributions from the left and right sides add together, resulting in a finite angular momentum during transport. Similar arguments hold for the term $-\hat{x}\hat{p}_y$. The same mechanism occurs when the bosons return from the outer region to the inner region. However, because the flow is now reversed, the angular momentum changes sign and becomes positive. Thus, the broken reflection symmetry of the potential converts the longitudinal flow into a finite oscillatory angular momentum.

The resulting angular-momentum dynamics closely follow the tunneling process, as evidenced by the nearly identical oscillation periods of the survival probability $\mathcal{P}_M(t)$ and the angular momentum.  As the particles move through the oppositely displaced squeezes, either from the middle region to the outer regions or back again, the directional flow generates a finite $\langle \hat{L}_Z\rangle$. Consequently, the largest values of $\vert \langle \hat{L}_Z\rangle \vert$ appear during the tunneling process. The angular momentum becomes smaller near the turning points of the tunneling cycle, where the survival probability $\mathcal{P}_M(t)$ approaches either $0$ or $1$. Furthermore, increasing $a$ makes this tunneling faster, as seen from the shorter time period of $\mathcal{P}_{M}(t)$. Therefore, the bosons passing through the squeezes produce a larger longitudinal-momentum contribution to $ \langle \hat{L}_Z\rangle$. As a result, larger $a$ leads to increased values of $\vert \langle \hat{L}_Z\rangle \vert$.

\subsection{Plurality of the collapse phenomenon}\label{manybody2}
\begin{figure}[htbp!]
    \centering
    \includegraphics[width=0.9\columnwidth]{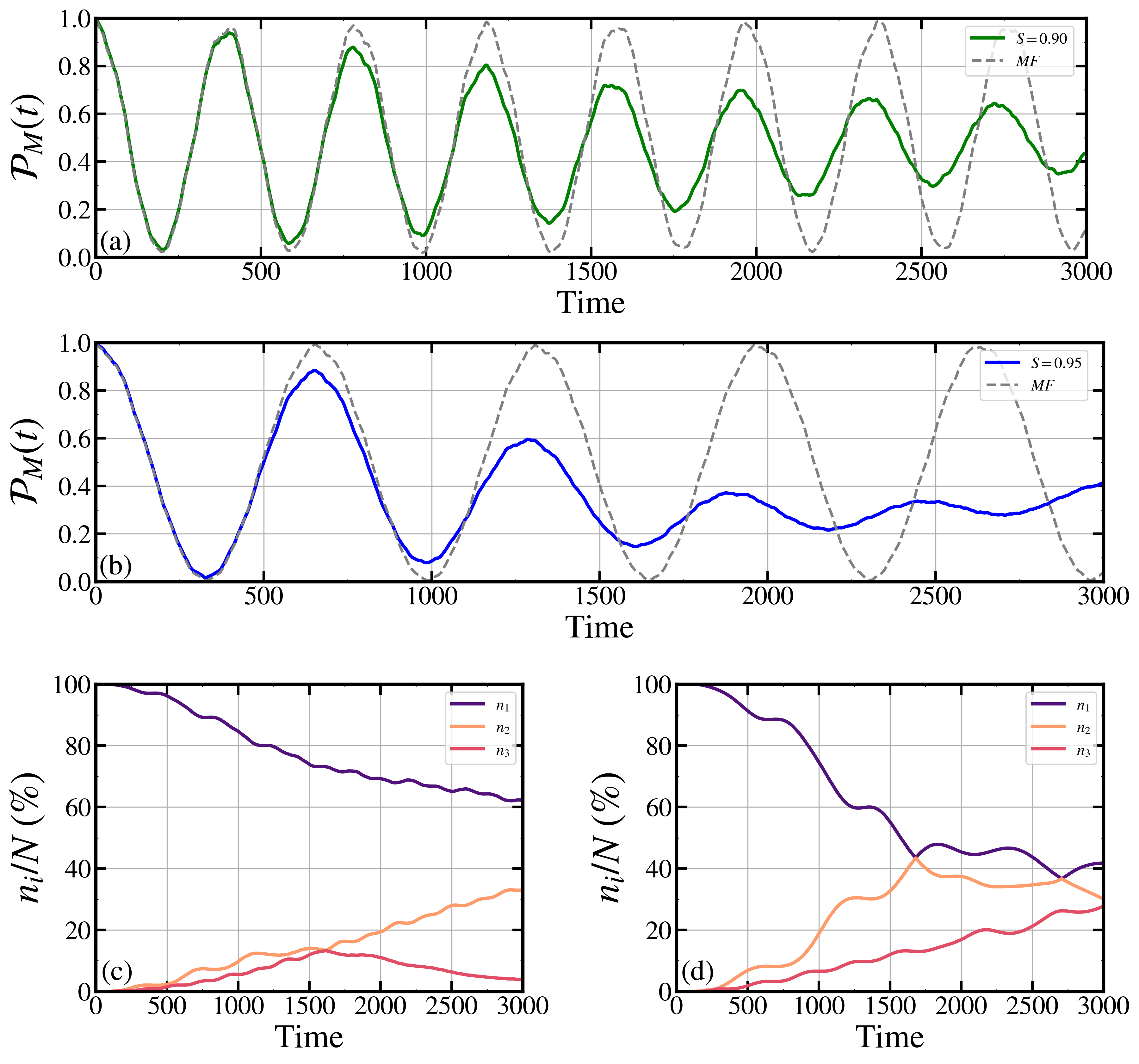}
    \caption{ Many-body dynamics in the modified squeezed channel with broken reflection symmetry, see Fig.~\ref{fig4}. (a,b) Survival probability $\mathcal{P}_{M}(t)$ for $S=0.90$ and $0.95$, respectively.  The solid curves show the many-body results (green and blue for $S=0.90$ and $S=0.95$, respectively), and the corresponding mean-field dynamics are shown by gray dashed curves. The two squeezes' magnitudes produce distinct collapse behaviors. (c,d) Occupations of the first three natural orbitals for (c) $S=0.90$ and (d) $S=0.95$. They exhibit different patterns of dynamical fragmentation. See the text for further discussions. All quantities shown are dimensionless.}
    \label{fig6}
\end{figure}

As discussed in Sec.~\ref{manybody1}, the many-body tunneling oscillations in the reflection-symmetric potential collapse around $50\%$ for different squeeze magnitudes. This raises the question of whether such a collapse is a general feature of the many-body dynamics or whether different collapse behaviors can be induced by modifying the trapping geometry. To address this question, we investigate the many-body dynamics in the modified squeezed potential, see Eqs.\eqref{eq:sq2}--\eqref{eq:squeezed_potential2} and Fig.~\ref{fig4}. Similar like the preceding many-body analysis of the reflection-symmetric squeezed potential, all other geometric parameters are kept fixed ($L=6.0$, $a=6.0$, $b=5.0$, $w=3.5$, $x_1=-4.0$, and $x_2=+4.0$), and the dynamics are examined for two squeeze magnitudes, $S=0.90$ and $S=0.95$. For each value of $S$, the longitudinal boundaries at $\pm x_0$ are adjusted based on the exact diagonalization analysis presented in Appendix~\ref{apndx1}, ensuring full tunneling at the mean-field level. The mean-field survival probabilities, shown by the dashed grey lines in Figs.~\ref{fig6}(a) and (b), exhibit full tunneling for both squeezes' magnitudes, with a longer oscillation period for $S=0.95$.

The many-body dynamics reveals distinct collapse behaviors for the two squeezes' magnitudes, as shown by the solid curves in Fig.~\ref{fig6}(a) and (b). The many-body dynamics initially follow the mean-field oscillations. As time evolves, their amplitudes gradually decay, leading to a collapse of the tunneling oscillations. This collapse of tunneling oscillations is associated with the dynamical buildup of fragmentation~\cite{sudip_atoms,anal_2020_dw}. But, it is important to note that the two squeezes' magnitudes lead to qualitatively different collapse patterns. For $S=0.90$, the oscillations collapse around $\mathcal{P}_{M}\approx 1/2$, resembling the reflection symmetric squeezed channel, where approximately half the population remains in the central region while the rest is distributed between the two outer regions. In contrast, increasing the squeeze magnitude to $S=0.95$ produces a fundamentally different collapse pattern, with the survival probability collapse to $\mathcal{P}_{M}\approx 1/3$. Hence, simply changing the magnitude of the squeeze can drive the system into a qualitatively different many-body tunneling regime, while the mean-field dynamics exhibit full tunneling in both cases. 

To understand the origin of this difference, we examine the  occupations in the three dominant natural orbitals, $n_1$, $n_2$ and $n_3$. The occupations in the lower three natural orbitals, $n_4$, $n_5$ and $n_6$, are presented in the supplemental material. Figures~\ref{fig6}(c) and \ref{fig6}(d) show $n_1$, $n_2$, and $n_3$ for $S=0.90$ and $S=0.95$, respectively. In both cases, the initial state is highly condensed, with $n_1\approx1$. During transport, the occupation of the first natural orbital decreases and both $n_2$ and $n_3$ starts to increase.  
For $S=0.90$, the occupation of the third natural orbital initially increases to about $n_3\sim 15\%$ and  then decreases, while $n_2$ continues to grow. Consequently, the first two natural orbitals remain dominant at longer times, with a smaller contribution from the third. This behavior accompanies the collapse of the survival probability oscillations to around $\mathcal{P}_{M}\approx 1/2$.
For $S=0.95$, the behavior of the occupation numbers is different. As $n_1$ decreases, the occupations of both the second and third natural orbitals increase. Unlike the $S=0.90$ case, the third natural orbital remains significantly occupied even at longer time. This distinct time evolution of the natural occupations accompanies the collapse of the survival probability oscillations to $\mathcal{P}_{M}\approx 1/3$. Thus, the the magnitude of the squeeze can control the buildup of fragmentation and lead to rich collapse dynamics.

In summary, tailoring the confinement offers a flexible way to control both quantum transport and correlated many-body dynamics in squeezed channels. By breaking the reflection symmetry of the potential, we demonstrate the generation of finite angular momentum and distinct many-body collapse behaviors, accompanied by different patterns of the dynamical fragmentation. A more complete understanding will require further exploration of the parameter regime, the role of interaction strength, and the relation between spatial transport and natural-orbital structure.

\section{Conclusions and outlook}\label{conclusion}
We have investigated how two localized squeezes in a channel potential influence the ground-state properties and dynamics of interacting bosons. These local squeezes generate effective longitudinal barriers that become stronger with increasing their magnitudes. 
As the squeeze magnitude increases, the ground-state density undergoes a gradual transition from a single delocalized cloud to three spatially localized clouds. Both the mean-field and many-body density distributions exhibit qualitatively similar behavior with increasing squeeze magnitude. Moreover, since this spatial redistribution is governed by the mean-field interaction parameter, the splitting of the density can be realized for different numbers of bosons. Despite the similarities in the density distributions, the many-body description reveals the emergence of fragmentation, a feature inaccessible at the mean-field level. At the many-body level, the change in the density distribution is accompanied by the emergence of many-body fragmentation, and strong squeezing eventually leads to a threefold fragmented ground state. Our results demonstrate that engineered channel geometry provides a flexible way to control both the spatial structure and the many-body properties of the ground state.

We then turned to the out-of-equilibrium dynamics in the reflection-symmetric two-squeeze geometry, focusing on the tunneling properties of the interacting bosons. Our results reveal that both the period of the tunneling oscillations and the degree of self-trapping can be controlled by tuning the shape of the potential, particularly the squeeze magnitude, the sharpness of the transverse confinement and the width of the channel. At the many-body level, the gradual buildup of fragmentation during the time evolution leads to the damping and eventual collapse of the density oscillations to around $50\%$. Despite the flexibility in controlling the transport, the expectation value of the angular momentum remains zero during the transport due to the reflection symmetry of the confinement. 

To extend this control further, we introduced a modified squeezed potential in which the two squeezes act from opposite transverse directions, thereby breaking the reflection symmetry of the confinement. In addition to retaining control over the tunneling dynamics, this geometrical modification enables the emergence of a nonvanishing, oscillatory angular momentum. Furthermore, the amplitude of the angular-momentum oscillations can be tuned by varying the confinement parameters. 
The modified geometry also exhibits distinct many-body collapse behaviors, accompanied by different patterns of dynamical fragmentation. In contrast to the reflection-symmetric geometry, where the density oscillations collapse to $50\%$, the modified geometry exhibits distinct collapse patterns depending on the shape of the confinement. This observation demonstrates how breaking the reflection symmetry and tailoring the shape of the confinement can influence the many-body dynamics and the associated fragmentation behavior. 

All in all, our findings demonstrate that squeezed channel potentials provide a versatile platform for controlling quantum transport, generating angular momentum, and influence many-body dynamics. Future studies could investigate how interactions and more complex squeezed geometries affect tunneling, angular-momentum dynamics, and many-body fragmentation, with the aim of designing geometry-controlled atomtronic and matter-wave transport devices.

\section*{Acknowledgements}
This work is supported by the Israel Science Foundation (Grant No. 2784/26). Computation time at the High-Performance Computing Center Stuttgart (HLRS) and High Performance Computing system Hive of the Faculty of Natural Sciences at the University of Haifa are gratefully acknowledged.

\appendix
\section{Exact diagonalization analysis of the squeezed potentials}\label{apndx1}

To obtain a microscopic understanding of the geometry-controlled transport, we perform the exact-diagonalization analysis of the pre- and post-quench single-particle Hamiltonians. The two-dimensional Hamiltonian is discretized on a uniform $128\times128$ grid using discrete variable representation (DVR). In this representation, the potential energy is represented by a diagonal matrix and the kinetic-energy matrix is constructed using the exponential DVR~\cite{3mctdh}. These two matrices are combined to form the total single-particle Hamiltonian matrix, which is then diagonalize to obtain the eigenvalues (energy levels) and eigenvectors (orbitals).

\begin{figure}[t]
    \centering
    \includegraphics[width=0.8\columnwidth]{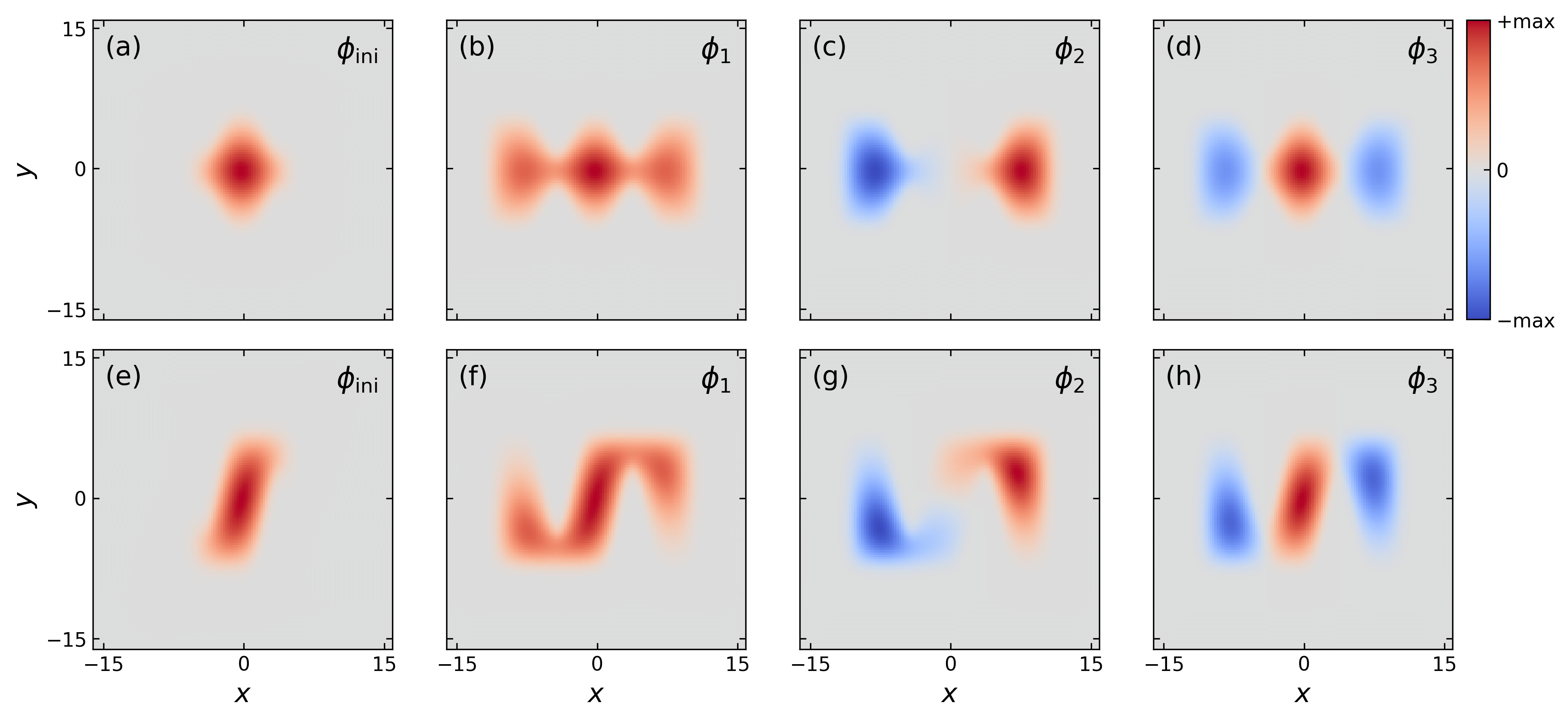}
    \caption{Shape of the orbitals obtained from exact diagonalization calculation. (a) Lowest eigenstate of the initial Hamiltonian for the symmetric geometry. (b)–(d) Three lowest eigenstates of the corresponding post-quench Hamiltonian. (e) Lowest eigenstate of the initial Hamiltonian for the modified geometry described in Sec.~\ref{results3}. (f)–(h) Three lowest eigenstates of the corresponding post-quench Hamiltonian.}
    \label{fig7}
\end{figure}

Figure~\ref{fig7} shows the orbitals obtained from the initial and post-quench Hamiltonians for the reflection-symmetric squeezed potential [Fig.~\ref{fig0}]. The initial state $\phi_{\text{ini}}$, localized in the central region, is shown in Fig.~\ref{fig7}(a). The three lowest eigenstates $\phi_1$, $\phi_2$, and $\phi_3$ of the post-quench Hamiltonian are shown in Figs.~\ref{fig7}(b)-(d). Due to the reflection symmetry of the potential, the initial state has nonzero overlaps with the first and third eigenstates of the post-quench Hamiltonian, and its overlap with $\phi_2$ vanishes. To quantify the contributions of the initial state to the post-quench eigenstates, we calculate the overlaps
\begin{equation}
O_i=\left|\langle \phi_{\text{ini}}|\phi_i\rangle\right|^2.
\end{equation}
We find that complete tunneling occurs when the initial state has approximately equal overlaps with $\phi_1$ and $\phi_3$, i.e., $O_1\approx O_3\approx 50\%$. To satisfy this condition, we adjust the longitudinal boundary position $x_0$, as illustrated in Fig.~\ref{fig0}, keeping all other parameters fixed. This analysis therefore helps us to choose the parameters needed to achieve complete tunneling.

The situation changes for the modified squeezed-channel geometry [Fig.~\ref{fig4}], shown in Figs.~\ref{fig7}(e)-(h). Here, the opposite squeezes at $x_1$ and $x_2$ modify  both the spatial structure and the localization properties of the  eigenstates. Consequently, complete transfer is no longer associated with equal contribution of the overlap from the first and third eigenstates. Instead, we find that complete tunneling occurs when the overlap ratio is $O_1: O_3\approx 60:40$. To satisfy this condition, we adjust the position $x_0$ while keeping all other parameters fixed. This overlap condition provides an empirical criterion for identifying the parameter regime that supports nearly complete tunneling in the modified squeezed potential. Thus, equal overlaps are not always necessary for complete tunneling, as the contributions of the eigenstates depend on their spatial distributions and the geometry of the confinement.

The single-particle overlap analysis provides insight into the tunneling mechanism and helps one identify parameter regimes favorable for complete transport. However, interparticle interactions can significantly affect the tunneling dynamics, giving rise to effects beyond the single-particle description. The full tunneling dynamics must therefore be studied by explicitly considering the interparticle interactions.

\bibliographystyle{apsrev4-2-title} 
\bibliography{zreferences}
\end{document}